\documentclass[10pt,twocolumn,twoside,journal]{IEEEtran}
\IEEEoverridecommandlockouts
\usepackage{subfigure} 
\usepackage{graphicx}

\usepackage{amsmath,graphicx,amssymb,mathtools,bm}
\usepackage{subfigure}
\usepackage{hyperref}
\usepackage{cite}
\usepackage{amsmath,amssymb,amsfonts}  
\usepackage{textcomp}
\usepackage{xcolor}
\usepackage{verbatim}  
\usepackage{bm}  
\usepackage{mathrsfs} 
 \usepackage{algorithmic} 
\usepackage{booktabs}
\usepackage{textcomp}  
\usepackage{multirow}  
\usepackage{lettrine}   
\usepackage{graphicx}  
\usepackage{color}  
\usepackage{amsmath}
\usepackage{amssymb}
\usepackage{stfloats} 
\usepackage{caption} 

\usepackage{color}  
 \DeclareMathSizes{12}{10.475}{7}{5}
\usepackage[lined,boxed,commentsnumbered]{algorithm2e}

\begin{document}
	\newcommand{\tabincell}[2]{\begin{tabular}{@{}#1@{}}#2\end{tabular}}
   \newtheorem{Property}{\it Property} 
  
 \newtheorem{Proposition}{\bf Proposition}
\newtheorem{remark}{Remark}
\newenvironment{Proof}{{\indent \it Proof:}}{\hfill $\blacksquare$\par}

\title{Flexible Precoding for Multi-User\\ Movable Antenna Communications 
}
 
\author{Songjie Yang, Wanting Lyu, Boyu Ning, \IEEEmembership{Member,~IEEE}, \\ Zhongpei Zhang, \IEEEmembership{Member,~IEEE}, and Chau Yuen, \IEEEmembership{Fellow,~IEEE}


\thanks{
	Songjie Yang, Wanting Lyu, Boyu Ning, and Zhongpei Zhang are with the National Key Laboratory of Wireless Communications, University of Electronic Science and Technology of China, Chengdu 611731, China. (e-mail:	yangsongjie@std.uestc.edu.cn; lyuwanting@yeat.net; boydning@outlook.com; zhangzp@uestc.edu.cn). \emph{(Corresponding Author: Zhongpei Zhang)}
	
		Chau Yuen is with the School of Electrical and Electronics Engineering, Nanyang Technological University (e-mail: chau.yuen@ntu.edu.sg).}}

\maketitle

\begin{abstract}
	This letter rethinks traditional precoding in multi-user wireless communications with movable antennas (MAs). Utilizing MAs for optimal antenna positioning, we introduce a sparse optimization (SO)-based approach focusing on regularized zero-forcing (RZF). This framework targets the optimization of antenna positions and the precoding matrix to minimize inter-user interference and transmit power. We propose an off-grid regularized least squares-based orthogonal matching pursuit (RLS-OMP) method for this purpose. Moreover, we provide deeper insights into antenna position optimization using RLS-OMP, viewed from a subspace projection angle. Overall, our proposed flexible precoding scheme demonstrates a sum rate that exceeds more than twice that of fixed antenna positions.

\end{abstract}
\begin{IEEEkeywords}
Movable antenna, flexible precoding, sparse optimization, regularized zero-forcing.
\end{IEEEkeywords} 
\section{Introduction}   
As wireless communication evolves, technological advancements in various dimensions are being continually made to achieve greater degrees-of-freedom (DoFs) and broader applications. One key area is the exploration of the wireless channel characteristic, harnessing its properties for enhanced communication and sensing. This includes utilizing the sparsity of millimeter-wave/terahertz channels for beamspace signal processing \cite{mmw1,mmw2}, employing reconfigurable intelligent surfaces to boost channel propagation \cite{RIS1,RIS2}, and exploiting near-field spherical-wave channels to unlock the distance DoF \cite{NF1,NF2}. Despite these developments, the potential of wireless channels remains an expansive field for further exploration.
Recently, movable antennas (MAs) have garnered significant interest for their ability to positively influence spatial channel paths through antenna movement \cite{MA4}, achieving effects like constructive/destructive path interference and path orthogonalization, collectively termed as modifying channel conditions. Similarly, fluid antenna systems (FASs) share a parallel concept with MAs, focusing on dynamic antenna position optimization \cite{FA1}.
Both MAs and FASs fundamentally align with the aforementioned techniques, as they all focus on harnessing the inherent capabilities of wireless channels.

Prior to the advent of MAs/FASs, two dynamic antenna position optimization techniques were extensively explored in the signal processing field. Antenna selection (AS) \cite{AS1}, involved finding the best antenna subset from a pre-defined dense array, effectively constituting discrete position optimization. The second technique, array synthesis \cite{ars1,ars2}, focused on achieving specific beam pattern objectives like sidelobe suppression, main lobe directivity enhancement, and efficient beam pattern representation using fewer elements, through the optimization of antenna positions, counts, and beamforming coefficients. Although differing in objectives and models, the methods employed in AS and array synthesis could provide valuable insights for MAs/FASs.

There have been several studies regarding performance analysis and signal processing for MAs and FAs \cite{FA1,FA2,FA3,MA1,MA2,MA3}. In \cite{MA1}, the authors maximized the multi-path channel gain under MA systems and showed the periodic behavior
of the multi-path channel gain in a given spatial field, providing insights for MA-enhanced communications.
The authors of \cite{FA1} derived a closed-form expression for the lower bound of capacity in FASs, confirming the substantial capacity gains derived from the diversity concealed within the compact space. Both \cite{MA1} and \cite{FA1} theoretically showcased the potential benefits of antenna position optimization in wireless communications. 
The study in \cite{FA4} explored point-to-point FAS communications using maximum ratio combining, revealing that the system's diversity order matches the total number of ports.
Moreover, \cite{FA2} noted in FASs that the multiplexing gain was directly proportional to the number of ports and inversely related to the signal-to-interference ratio target. Meanwhile, \cite{MA2,FA3} showed the potential of uplink power minimization through optimizing the user's antenna position. Conversely, \cite{MA3} explored MAs at the base station (BS) end, utilizing particle swarm optimization to optimize antenna positions with the objective of maximizing the minimum user rate. However, such meta-heuristic algorithms, despite their effectiveness, often come with high computational complexity and may lack deeper theoretical insights.

To grasp how MAs enhance multi-user communications, we concentrate on traditional precoding schemes and examines the modifications introduced by MAs. Therefore,
\textbf{\emph{what will happened when traditional precoding meets MAs?}} Traditional precoding methods show effectiveness under various channel conditions, yet they fall short of ideal performance, mainly due to challenges like multipath effects and inter-user coherence.  To overcome these issues, we enhance traditional precoding with MAs, thus introducing flexible precoding. 

 Our focus is on regularized zero-forcing (RZF) precoding, which can be adapted into maximum ratio transmission (MRT), ZF, or minimum mean square error (MMSE) precoding by tuning the regularization factor.  We are innovating RZF into a more flexible version by optimizing antenna positions, driven by two primary perspectives and contributions\footnote{\label{note1}The source code of this work is open in \url{https://github.com/YyangSJ/Flexible_Precoding_Movable_Antennas} for readers studying.}:
\begin{itemize}
	\item \textbf{Sparse Optimization (SO):} 
	Given the RZF precoding problem, along with the nature of MAs that optimize antenna positions within a movable region, we propose a flexible RZF formulation as a SO framework.
	This framework employs an on-grid dictionary representation for antenna positions, complemented by off-grid (OffG) optimization techniques.
However, the SO problem is uncommon and challenging to solve due to the minimal inter-element spacing constraint. To address this, we propose an OffG regularized least squares-based orthogonal matching pursuit (RLS-OMP), specifically crafted to efficiently resolve the SO problem.
	
	\item \textbf{Subspace Projection:} By leveraging the relationship between column-subspace projection and RLS, we dissect the core mechanism of OffG RLS-OMP to elucidate its role in antenna position optimization. This reveals that optimizing the positions of $N$ antennas is essentially about pinpointing an optimal subspace spanned by the array-position manifolds at these specific $N$ locations. 
\end{itemize}

{\emph {Notations}}:
  ${\left(  \cdot  \right)}^{ T}$, ${\left(  \cdot  \right)}^{ H}$, and $\left(\cdot\right)^{-1}$ denote   transpose, conjugate transpose, and inverse, respectively. $\vert\cdot\vert$ denotes the cardinality of a set.
$\Vert\cdot\Vert_0$ and $\Vert\cdot\Vert_1$ represent $\ell_0$ norm and $\ell_1$ norm, respectively. 
$\Vert\mathbf{A}\Vert_F$ denotes the Frobenius norm of matrix $\mathbf{A}$.     $[\mathbf{A}]_{i,:}$ and $[\mathbf{A}]_{:,j}$ denote the $i$-the row and the $j$-the column of matrix $\mathbf{A}$, respectively.  $\mathbb{E}\{\cdot\}$ denotes the expectation.  
\vspace{-0.33cm}

\section{System Model}

We consider a MU-MISO downlink system, where the BS is equipped with $N\triangleq N_x\times N_z$ movable antenna elements, distributed along the $x$-$z$ plane, while $K$ users are equipped with a single fixed antenna. The BS supports a single data stream for each user.

In the downlink communication system, the received signal at the $k$-th user, $k\in\{1,\cdots,K\}$, can be expressed by
\begin{equation}
	y_k=\mathbf{h}_k^H\mathbf{F}\mathbf{s}+n_k,
\end{equation}
where $\mathbf{h}_k\in\mathbb{C}^{N\times 1}$ is the $k$-th user's channel, $\mathbf{F}\triangleq[\mathbf{f}_1,\cdots,\mathbf{f}_K]\in\mathbb{C}^{N\times K}$ is the precoding matrix,
$\mathbf{s}\in\mathbb{C}^{K\times 1}$ represents the $K$ data streams for $K$ users, $n_k$ represents the Gaussian additive white noise following $\mathcal{CN}(0,\sigma^2)$.

By assuming the i.i.d. transmit data such that $\mathbb{E}\{s_k^*s_k\}=1$ and $\mathbb{E}\{s_i^*s_k\}=0$, $\forall k , k\neq i$,
the SINR of the $k$-user can be given by
\begin{equation}
	{\rm SINR}_k=\frac{\vert\mathbf{h}_k^H\mathbf{f}_k\vert^2}{\sum_{i,i\neq k}^{K}\vert\mathbf{h}_k^H\mathbf{f}_i\vert^2+\sigma^2}.
\end{equation} 

 The significance of MASs relies on the spatial channel model, by assuming $L$ paths for all user channels, the channel is given by, for $\forall k$,
 \begin{equation}\label{hk}
 	\mathbf{h}_k=\sqrt{\frac{1}{L}}\sum_{l=1}^{L}\beta_{k,l} \mathbf{a}(\theta_{k,l},\phi_{k,l}),
 \end{equation}
where $\beta_{k,l}$ is the complex path gain of the $l$-th path of the $k$-user's channel, $L$ is number of spatial channel paths,
$\phi_{k,l}\triangleq \sin(\varphi_{k,l})\sin(\vartheta_{k,l})$, and $\theta_{k,l}\triangleq\cos(\vartheta_{k,l})$ denote the virtual DoAs of the $l$-the path of the $k$-th user channel, and the array-angle manifold $\mathbf{a}(\theta_{k,l},\phi_{k,l})\in\mathbb{C}^{N\times 1}$ follows
\begin{equation}
	\begin{aligned}
		\mathbf{a}(\theta_{k,l},\phi_{k,l})&=\left[ e^{j\frac{2\pi}{\lambda}(\phi_{k,l} x_1+\theta z_1)},\cdots, \right. \\ &  \left. e^{j\frac{2\pi}{\lambda}(\phi_{k,l} x_{n_x}+\theta z_{n_z})},\cdots, e^{j\frac{2\pi}{\lambda}(\phi_{k,l} x_{N_x}+\theta_{k,l} z_{N_z})} \right]^T,
	\end{aligned}
\end{equation}
where $(x_{n_x},z_{n_z})$ is the position of the $(n_x,n_z)$-th antenna in the $x$-$z$ plane.
Noticing that the array normalization factor $1/\sqrt{N}$ is removed by ignoring the array gain in Eqn. (\ref{hk}).
\vspace{-0.33cm}
\section{SO-Based Flexible RZF Precoding}
By stacking the channel vectors of all users in rows, we obtain $\mathbf{H}=[\mathbf{h}_1,\cdots,\mathbf{h}_K]^H\in\mathbb{C}^{K\times N}$. Following this, the linear precoding scheme, RZF, addresses the problem as follows:
\begin{equation}\label{FZF_pro}
	\underset{\mathbf{F}}{{\rm arg \ min}} \	\left\Vert\mathbf{I}_K-\mathbf{H}\mathbf{F}\right\Vert_F^2+\alpha\left\Vert \mathbf{F} \right\Vert_F^2.
	\tag{$\mathcal{P}_1$}
\end{equation}
where $\alpha$ is the regularization factor. Its solution is
\begin{equation}\label{RZF}
	\mathbf{F}=\mathbf{H}^H (\mathbf{H}\mathbf{H}^H+\alpha\mathbf{I} )^{-1}.
\end{equation}

Let $\alpha$ be adjustable, RZF is equivalent to MRT, ZF, and MMSE precoding:
\begin{equation}
\mathbf{F}=	\begin{cases}
		\mathbf{F}_{\rm MRT}\triangleq \mathbf{H}^H, \alpha\rightarrow\infty; \\
	\mathbf{F}_{\rm ZF}\triangleq \mathbf{H}^H (\mathbf{H}\mathbf{H}^H)^{-1},\alpha = 0; \\
		\mathbf{F}_{\rm MMSE}\triangleq \mathbf{H}^H (\mathbf{H}\mathbf{H}^H+\sigma^2\mathbf{I} )^{-1}, \alpha =\sigma^2.
	\end{cases}
\end{equation}

Consequently, our analysis focuses on RZF precoding to glean insights applicable to other precoding schemes. 
 

In MASs, the variables in problem (\ref{FZF_pro}) extend beyond just precoding to also include the channel. Instead of optimizing the channel matrix $\mathbf{H}$ directly,  $\{\mathbf{x},\mathbf{z}\}$, representing the $x$- and $z$-axis coordinates of the $N$ antennas, are optimized. Consequently, this adjustment allows us to establish a new problem formulation that incorporates constraints on the antenna positions. 
\begin{equation}\label{FZF_2}
	\begin{aligned}
		\underset{\mathbf{F},\mathbf{x},\mathbf{z}}{{\rm arg \ min}}& \	\left\Vert\mathbf{I}_K-\mathbf{H}\mathbf{F}\right\Vert_F^2+\alpha\left\Vert \mathbf{F} \right\Vert_F^2\\
		 {\rm s.t.} \ & \sqrt{(x_i-x_{j})^2+(z_i-z_j)^2}\geq \lambda/2, \\
		 & \forall i,j\in\{1,\cdots,N\}, \ i\neq j, \\
	& (\mathbf{x},\mathbf{z})\in \bm{\mathcal{A}},
	\end{aligned}
	\tag{$\mathcal{P}_2$}
\end{equation}
where $(\mathbf{x},\mathbf{z})$ represents all antenna coordinates, which are limited in the two-dimensional continous feasible region $\bm{\mathcal{A}}$.

 
Acknowledging the linear regression structure of problem (\ref{FZF_2}), and understanding that optimizing antenna positions is akin to locating $N$ antennas within a continuous feasible region (essentially selecting $N$ from an infinite/huge set), we are inspired to formulate the subsequent SO problem. This falls under the OffG or gridless SO frameworks, aiming to achieve a super-resolution solution for the parameters:
\begin{equation}\label{SO}
	\begin{aligned}
		\underset{\widetilde{\mathbf{F}},\mathbf{x},\mathbf{z}}{{\rm arg \ min}} \ &	\left\Vert\mathbf{I}_K-\widetilde{\mathbf{H}}\widetilde{\mathbf{F}}\right\Vert_F^2+\alpha\left\Vert \widetilde{\mathbf{F}} \right\Vert_F^2 \\
		{\rm s.t.} 	& \left\Vert \widetilde{\mathbf{F}}\right\Vert_{0,\rm row}= N, \\ &\sqrt{(x_i-x_{j})^2+(z_i-z_j)^2}\geq \lambda/2, \\
		& \forall i,j\in\{1,\cdots,N\}, \ i\neq j, 
	\end{aligned}
	\tag{$\mathcal{P}_3$}
\end{equation}
where $\left\Vert\cdot\right\Vert_{0,\rm row}$ denotes the number of non-zero rows in an matrix, and $\widetilde{\mathbf{H}}\in\mathbb{C}^{K\times G}$ symbolizes the virtual channel or `dictionary' with $G$ representing $G\triangleq G_xG_z$ feasible positions defined by  $\bm{\mathcal{{G}}}=\{(x,z)|x=0,d,2d,\cdots,(G_x-1)d; z=0,d,2d,\cdots,(G_z-1)d\}$ with $d\triangleq \lambda/2$.
$\bm{\mathcal{G}}$ is the discretized counterpart of $\bm{\mathcal{A}}$ in problem (\ref{FZF_2}).
 Each column of $\widetilde{\mathbf{H}}$ adheres to an array-position manifold:
\begin{equation}
	\begin{aligned}
	&	\mathbf{b}(x,z)=\frac{1}{\sqrt{L}}\left[\sum_{l=1}^L \beta_{1,l}^* e^{-j\frac{2\pi}{\lambda} (\phi_{1,l}x+\theta_{1,l}z) },\cdots, \right. \\ & \left.
		\sum_{l=1}^L \beta_{k,l}^* e^{-j\frac{2\pi}{\lambda} (\phi_{k,l}x+\theta_{k,l}z) },\cdots,	\sum_{l=1}^L \beta_{K,l}^* e^{-j\frac{2\pi}{\lambda} (\phi_{K,l}x+\theta_{K,l}z) }
\right]^T.
	\end{aligned}
\end{equation}

This suggests that the virtual channel $\widetilde{\mathbf{H}}$ can be expressed as $[\mathbf{b}(x_1,z_1),\mathbf{b}(x_2,z_1),\cdots,\mathbf{b}(x_{G_x},z_{G_z})]$. Consequently, problem (\ref{SO}) is focused on optimizing both the antenna positions and precoding, using this virtual channel as a dictionary. This optimization incorporates OffG refinement or other super-resolution techniques to determine continuous positions.

 When we ignore the regularization term $\alpha\left\Vert \widetilde{\mathbf{F}} \right\Vert_F^2$ and the constraints on $\mathbf{x}$ and $\mathbf{z}$, the problem (\ref{SO}) becomes a conventional OffG SO problem. However, incorporating these components renders this particular SO problem rather distinctive.
 
 The proposed method, OffG RLS-OMP, utilizes an $N$-iteration process. In each iteration, it identifies and optimizes the most suitable antenna position, then removing its effect for next iteration. 
Additionally, in Section \ref{insi}, we provide insights from a subspace projection viewpoint, addressing \textbf{\emph{why optimizing antenna positions in an OMP-like manner is feasible?}}

The comprehensive workflow of flexible RZF precoding is detailed in Algorithm \ref{FP}, encompassing the following steps:
\begin{itemize}

\item \emph{Antenna Matching:}
Given the residual matrix $\mathbf{R} \in \mathbb{C}^{K \times K}$, which initially begins as $\mathbf{I}_K$, our goal is to find an antenna that maximizes the matching gain for all users. This method efficiently identifies key antennas. The selection is mathematically represented as:
$g^\star=\underset{g\in\{1,\cdots,G\}}{{\rm arg \ max}} \ \left\Vert \  \left[\overline{\mathbf{H}} \right]_{:,g} ^{H} {\mathbf{R}}  \right\Vert_1$.

\item \emph{Position Optimization:}
In the \emph{Antenna Matching} stage, it is observed that the antenna is positioned on a grid $\mathcal{G}$, allowing for OffG refinement to enhance DoFs. This step, termed $\textbf{OffG\_Position\_Optimization}(\cdot)$, is elaborated in Appendix \ref{OFFG}.
 
 \item \emph{RLS:} Leveraging the channel matrix $\mathbf{H}^\star$, shaped by the chosen and fine-tuned antenna positions, we then proceed to compute the their $n$-th iteration coefficients with RLS
 \begin{equation}
 	\mathbf{F}^\star =\left(\mathbf{H}^{\star,H}\mathbf{H}^\star+\alpha \mathbf{I}_n\right)^{-1}\mathbf{H}^{\star,H}
 \end{equation}
\item \emph{Residual Calculation:} After obtaining the precoding based on optimized antenna positions, the residual is determined by cancelling the existing antennas' impact, calculated by 
$\mathbf{R}=\mathbf{I}_K-\mathbf{H}^{\star}\mathbf{F}^\star$. 
	\item \emph{Support Confirmation:} 
Before proceeding with antenna matching for next iteration, it is critical to confirm that the distance variation between each antenna in the support set and their optimized positions exceeds the minimum allowable spacing. Any antennas failing to meet this requirement must be excluded from the support set.
\end{itemize}
\begin{algorithm} 
	\caption{Flexible Precoding via OffG RLS-OMP.} 
	\label{FP}
	\KwData {$N$, $K$, $\bm{\theta}$, $\bm{\phi}$,  $\overline{\mathbf{H}}$, and $\alpha$.	}
	\KwResult {Optimized position set $\widehat{\bm{\mathcal{G}}}$ and precoding $\mathbf{F}^\star$.}
	\BlankLine
	\Begin{ 
		$\textbf{Initialization:}$ $\mathbf{R}=\mathbf{I}_K$, $\widehat{\bm{\mathcal{G}}}=\emptyset$, $\bm{\Gamma}=\{1,\cdots,G\}$.\\
		\For{$n=1,\cdots,N$}{ 
			\tcp{\emph{Antenna Matching}} 	  $g^\star=\underset{g\in\bm{\Gamma}}{{\rm arg \ max}} \ \left\Vert \  \left[\overline{\mathbf{H}} \right]_{:,g} ^{H} {\mathbf{R}}  \right\Vert_1$ \;
			$g^\star\Rightarrow (x_n^\star,z_n^\star)$, \
			$ \widehat{\bm{\mathcal{G}}}\leftarrow  \widehat{\bm{\mathcal{G}}}\cup(x_n^\star,z_n^\star)$\;  \tcp{\emph{Position Optimization}}
			$\widehat{\bm{\mathcal{G}}}=\textbf{OffG\_Position\_Refinement} \left(
			\widehat{\bm{\mathcal{G}}},\bm{\theta},\bm{\phi}\right)$\;  \tcp{\emph{Coefficient Calculation}} 
			$\mathbf{H}^\star=[\mathbf{b}(x_1^\star,z_1^\star),\cdots,\mathbf{b}(x_n^\star,z_n^\star)]$ \;
			$\mathbf{F}^\star=\left(\mathbf{H}^{\star,H}\mathbf{H}^\star+\alpha \mathbf{I}_n\right)^{-1}\mathbf{H}^{\star,H}$ \;	\tcp{\emph{Residual Updating}} 	 	$\mathbf{R}=\mathbf{I}_K-\mathbf{H}^\star\mathbf{F}^\star$\;
			\tcp{\emph{Support Confirmation}} 	 \If{$\sqrt{(x_i-x_n^\star)^2+(z_i-z_n^\star)^2}>\frac{\lambda}{2}$, $\forall i\in\bm{\Gamma}$}{$\bm{\Gamma}\leftarrow \bm{\Gamma}\setminus  i$}  }  
		Normalize each column of $\mathbf{F}^\star$.
	}	\Return{$\widehat{\bm{\mathcal{G}}}$ and $\mathbf{F}^\star$}
\end{algorithm}
 \begin{figure*}
 	\centering 
 	\includegraphics[width=6.83in]{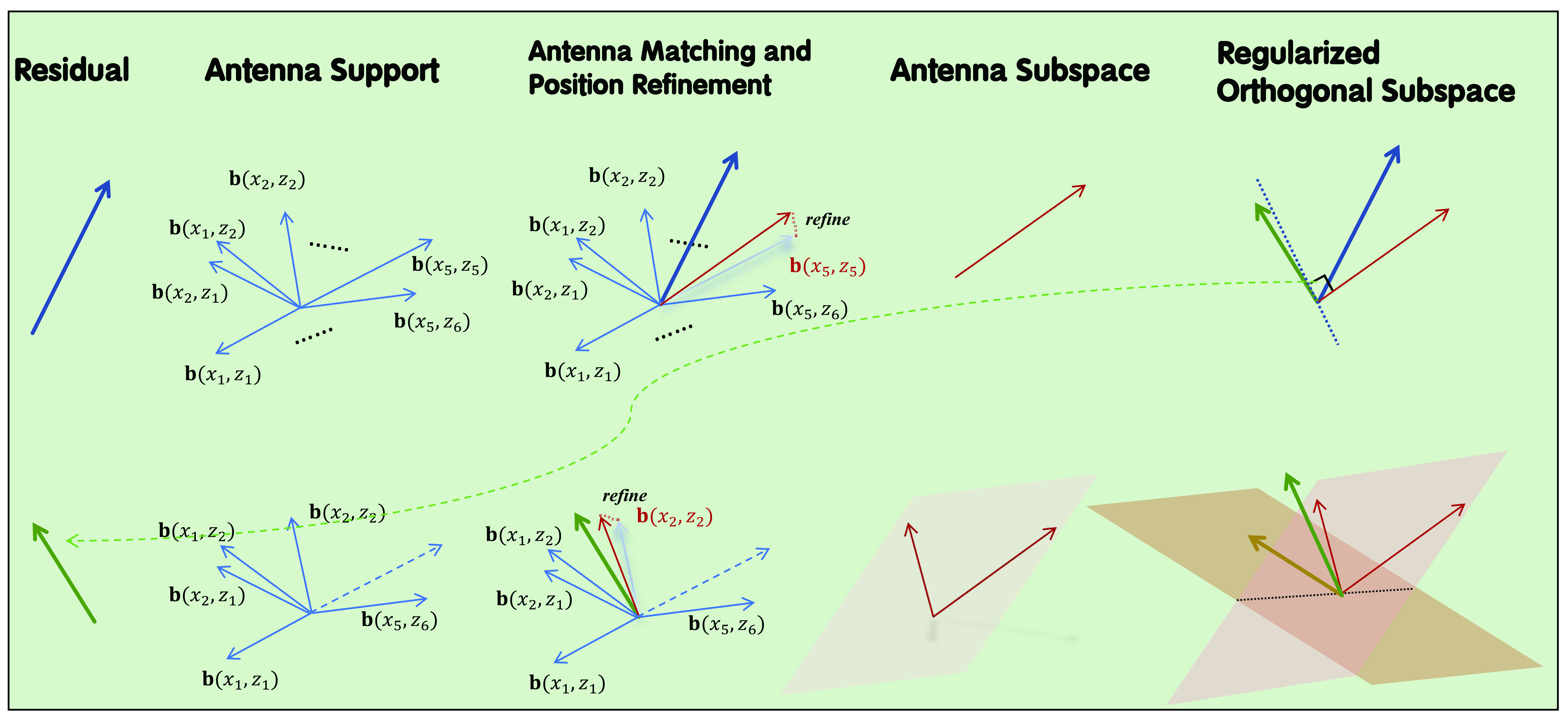}
 	\caption{The two-iteration antenna position optimization from the perspective of subspace projection.}\label{SP} 
 \end{figure*}  
\section{Insights Into Flexible RZF Precoding}\label{insi}
\vspace{-0.16cm}
Last section proposes the implementation of flexible RZF precoding, adapted from OMP within the context of sparse optimization. This section delves deeper, analyzing why the proposed OffG RLS-OMP's greedy approach is effective in achieving flexible precoding.

The expression of RZF precoding in Eqn. (\ref{RZF}) is equivalent to the RLS in Algorithm \ref{FP} for $\alpha\neq 0$:
 \begin{equation}\label{RZF2}
 	\begin{aligned}
 		\mathbf{F}&=\mathbf{H}^H (\mathbf{H}\mathbf{H}^H+\alpha\mathbf{I} )^{-1} \\ &= (\mathbf{H}^H\mathbf{H}+\alpha\mathbf{I} )^{-1}\mathbf{H}^H.
 	\end{aligned}
 \end{equation}

It should be noted that this equation does not hold at $\alpha=0$ but it hold when $\alpha\rightarrow 0$. Therefore, if we want to use Eqn. (\ref{RZF2}) for ZF precoding,  $\alpha\rightarrow 0$ instead of $\alpha =0$ should be satisfied. Also now, noticing that RZF aligns with the coefficient calculation step in Algorithm \ref{FP}, which indicates in each iteration of Algorithm \ref{FP}, we will calculate RZF precoding according to the present optimized channel.

The matrix $\mathbf{H}(\mathbf{H}^H\mathbf{H})^{-1}\mathbf{H}^H$ is commonly understood as the projection onto the subspace $\bm{\mathcal{H}}$, spanned by the columns of $\mathbf{H}$. In a similar vein, $\mathbf{H}(\mathbf{H}^H\mathbf{H}+\alpha\mathbf{I})^{-1}\mathbf{H}^H$ acts as a regularized projection onto $\bm{\mathcal{H}}$, albeit with directional adjustments\footnote{Strictly speaking, it is not a projection matrix, as it does not fulfill $\mathbf{P}^2=\mathbf{P}$, where $\mathbf{P}\triangleq\mathbf{H}(\mathbf{H}^H\mathbf{H}+\alpha\mathbf{I})^{-1}\mathbf{H}^H$.}. This approach effectively `shrinks' the projection, proving advantageous in addressing numerical ill-conditioning and noise-related issues. Correspondingly, $\mathbf{I}-\mathbf{H}(\mathbf{H}^H\mathbf{H}+\alpha\mathbf{I})^{-1}\mathbf{H}^H$ serves as a regularized orthogonal complement and aligns with the \emph{Residual Updating} step in Algorithm \ref{FP}.

 In MAs, the subspace $\bm{\mathcal{H}}$ can be dynamically spanned through adjusting antenna positions, offering the potential to identify an optimal subspace for precoding in different channel conditions. Consequently, the goal of flexible precoding becomes finding the most suitable subspace $\bm{\mathcal{H}}^\star$ for projecting $\mathbf{I}_K$. We now explore how the proposed OffG RLS-OMP method meets this objective.
 
 To elucidate, let us start by considering the projection of a vector, such as the first column of $\mathbf{I}_K$, denoted as $\mathbf{i}_1$, in a quest to attain $\bm{\mathcal{H}}^\star$ using a greedy approach. In the initial iteration, we compute the inner product of $\mathbf{i}_1$ with the array-position manifolds from the antenna support set, selecting the optimal antenna and subsequently refining its position. This aligns with the \emph{Antenna Matching} and \emph{Position Optimization} steps. Following this, we establish the antenna subspace and its regularized projection depending all columns of $\mathbf{I}_K$, which then informs the residual calculation for the subsequent iteration. The two-step process is illustrated in Fig. \ref{SP}. Ultimately, the aim is to pinpoint the optimal antenna subspace conducive to precoding.

\section{Simulation Results} 

The system operates at a central frequency of $3$ GHz. The BS is equipped with $N=4$ antennas, serving $K=4$ users. Both the total transmit power and the noise power are standardized to $1$. The user and scatterer locations are uniformly distributed with azimuth angles $\phi_{k,l}$ and elevation angles $\theta_{k,l}$ ranging within $[-1,1]$ for all $k,l$. All users are assumed to have an identical number of channel paths and equal power.
 Key methods included in this evaluation are: 1) Fixed Antenna Position, applying RZF precoding with $N$ antennas at half-wavelength spacing; 2) Fast AS, combining RZF precoding with AS in \cite{AS1}; and 3) Flexible Precoding, incorporating our proposed approach detailed in Algorithm \ref{FP}.

\begin{figure*} 
 	\centering
 	\begin{minipage}[b]{0.323\linewidth} 
 		\centering
 		\includegraphics[width=\linewidth]{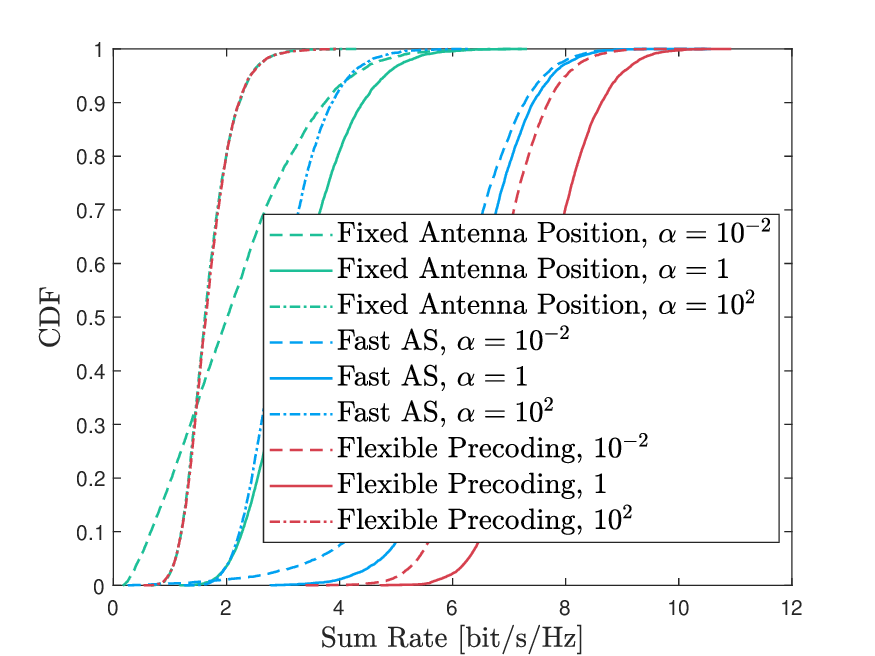}
 		\caption{CDF versus sum rate.}
 		 \label{CDF}
 	\end{minipage}
 	\begin{minipage}[b]{0.323\linewidth} 
 		\centering
 		\includegraphics[width=\linewidth]{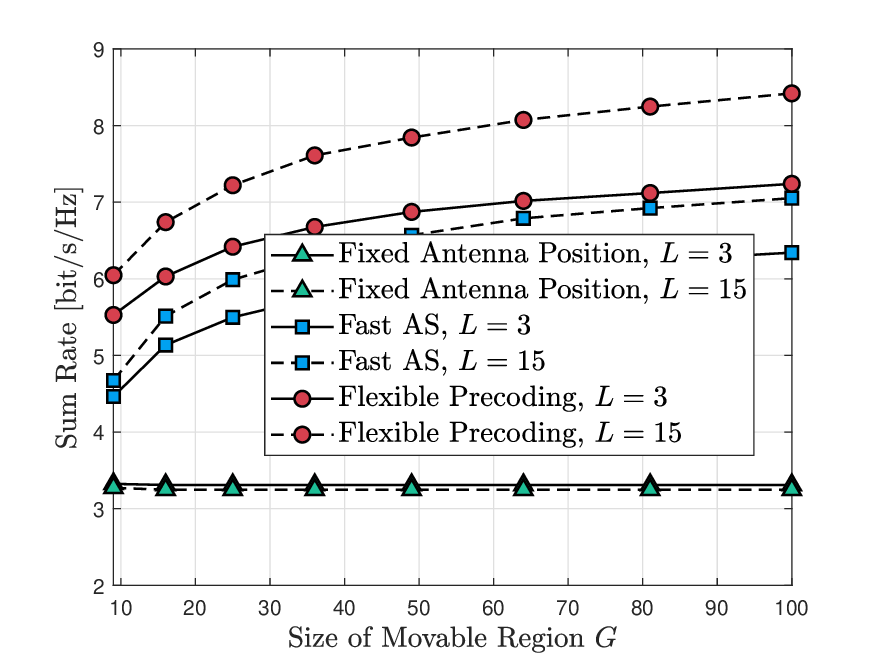}
 		\caption{Sum rate versus $G$.}
 		\label{SG}
 	\end{minipage}
 	\begin{minipage}[b]{0.323\linewidth} 
 		\centering
 		\includegraphics[width=\linewidth]{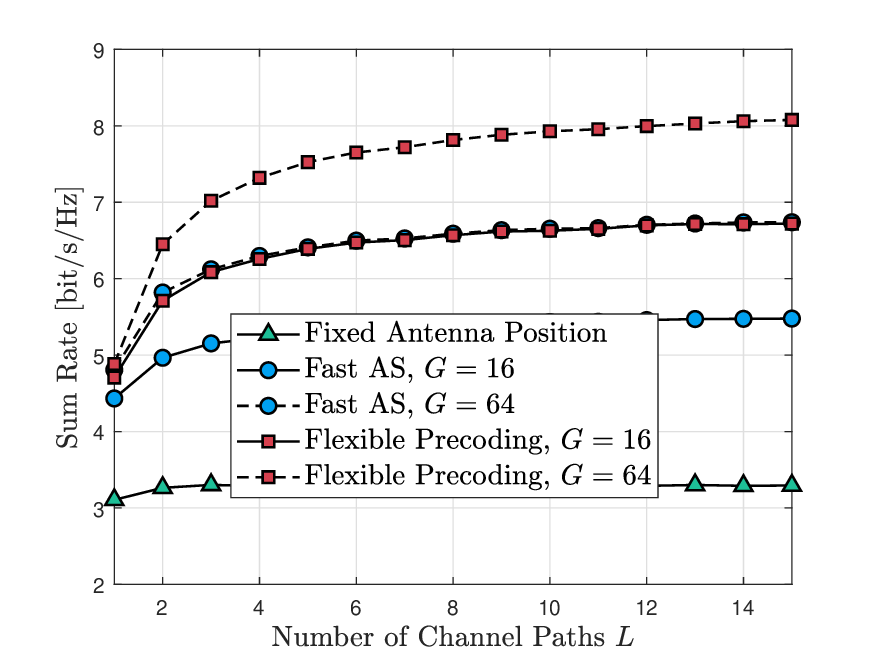}
 		\caption{Sum rate versus $L$.}
 		\label{SL}
 	\end{minipage} 
 	\label{M_L}
 \end{figure*}

The first simulation assesses the CDF for sum rate across the three methods when the movable region $G=36$, the number of paths $L=15$, and the $\alpha$ values are set to $10^{-2},1,10^{2}$ for ZF, MMSE, and MRT, respectively. As shown in Fig. \ref{CDF}, flexible precoding for ZF and MMSE demonstrates superior performance compared to the other methods. For MRT, flexible precoding achieves similar results to the fixed approach but falls short of fast AS. 

Fig. \ref{SG} presents the sum rate performance of different methods when $\alpha=1$, $L\in\{3,15\}$, and the movable region size $G$ is in $\{3^2,4^2,5^2,6^2,7^2,8^2,9^2,10^2\}$. An upward trend with increasing $G$ is apparent for both fast AS and flexible precoding, with a more pronounced increase for $L=15$ compared to $L=3$. Both fast AS and flexible precoding demonstrate the benefits of antenna position adjustment. Particularly, when flexible precoding is within a small range of movement, such as $G=9$, its performance is roughly twice that of fixed antenna position, and as $G$ increases, the performance gap between them becomes larger. 

In Fig. \ref{SL}, the effect of the number of channel paths on sum rate is investigated, with $G=36$ and $\alpha=1$. As $L$ grows, both fast AS and flexible precoding exhibit improvements. Echoing the observations from Fig. \ref{SG}, flexible precoding excels beyond fast AS and delivers performance more than double that of fixed antenna positions when both $L$ and $G$ are large.
  
\section{Conclusions}\label{Con} 

This letter presents a groundbreaking framework for optimizing antenna positions in MAs, particularly focusing on flexible RZF precoding through SO. Utilizing the  proposed OffG RLS-OMP algorithm, it efficiently addresses  antenna position optimization under minimal inter-element spacing constraints and flexible precoding. 
The essence of this flexible precoding strategy is further elucidated through the concept of subspace projection, highlighting the core objective of optimizing antenna positions as finding the most effective antenna subspace.
The novel strategy we propose offers new insights for MAs and has been shown to significantly outperform traditional fixed antennas and the classical AS method. Beyond the RZF precoding discussed in this letter, numerous other applications in MAs can be explored using our proposed approach.
 
\begin{appendices}
\section{Derivation of OffG Position Refinement}\label{OFFG}
 Given the antenna position obtained in the $n$ iteration of Algorithm \ref{FP}: $(x_n,z_n)$, the optimized position set $\widehat{\bm{\mathcal{G}}}$, and the DoAs $\{\bm{\theta},\bm{\phi}\}$, we can construct the channel by $\mathbf{H}=[\mathbf{b}(x_1,z_1),\cdots,\mathbf{b}(x_n,z_n)]$. We aim to optimize $(x_n,z_n)$ in the $n$ iteration, with
 \begin{equation}
 	\begin{aligned}
 	&	\underset{{\mathbf{F}},x_n,z_n}{{\rm arg \ min}} \ 	\left\Vert\mathbf{I}-{\mathbf{H}}{\mathbf{F}}\right\Vert_F^2+\alpha\left\Vert {\mathbf{F}} \right\Vert_F^2 \\
 		{\rm s.t.} \ & \sqrt{(x_n-x_{t})^2+(z_n-z_t)^2}\geq \lambda/2, \\
 		& \forall t\in\{1,\cdots,n-1\}.
 	\end{aligned}
 	\tag{$\mathcal{P}_{3.1}$}
 \end{equation}

By adding permutations on $x_n
 \leftarrow x_n+\eta_n $ and $z_n \leftarrow z_n+\xi_n$ , and using first-order Taylor expansion, yielding
 \begin{equation}
 	\begin{aligned}
 		\mathbf{b}(x_n+\eta_n ,z_n+\xi_n)\approx & \mathbf{b}(x_n,z_n) + \eta_n\frac{\partial \mathbf{b}(x_n, z_n)}{\partial x_n}+ \\& \xi_n\frac{\partial \mathbf{b}(x_n, z_n)}{\partial z_n}.
 	\end{aligned}
 \end{equation} 
 
 
 The above statement indicates that the perturbation optimization is a form of fine-tuning optimization:
 \begin{equation}\label{TR}
 	\begin{aligned}
 		\underset{{\eta}_n,{\xi}_n}{{\rm arg \ min}} \ &	\left\Vert\mathbf{R}-	\left(\eta_n\frac{\partial \mathbf{b}(x_n, z_n)}{\partial x_n}+ \xi_n\frac{\partial \mathbf{b}(x_n, z_n)}{\partial z_n}\right)\mathbf{f}_n\right\Vert_F^2 \\
 {\rm s.t.} \ & \sqrt{(x_n-x_{t})^2+(z_n-z_t)^2}\geq \lambda/2, \\
 & \forall t\in\{1,\cdots,n-1\},
 	\end{aligned}
 	\tag{$\mathcal{P}_{3.2}$}
 \end{equation}
 where $\mathbf{f}_n\triangleq [\mathbf{F}]_{n,:}$, and $\mathbf{R}\triangleq \mathbf{I}-\mathbf{HF}$ is the residual.
 
 Denoted by the objective $\mathcal{U}\triangleq \mathbf{R}-	\left(\eta_n\frac{\partial \mathbf{b}(x_n, z_n)}{\partial x_n}+ \xi_n\frac{\partial \mathbf{b}(x_n, z_n)}{\partial z_n}\right)\mathbf{f}_n$, we re-write it in vectors:
 \begin{equation}
 	\begin{aligned}
 		{\rm vec}(\mathcal{U})=&\mathbf{r}- \mathbf{f}_n^T \otimes 	\left(\eta_n\frac{\partial \mathbf{b}(x_n, z_n)}{\partial x_n}+ \xi_n\frac{\partial \mathbf{b}(x_n, z_n)}{\partial z_n}\right)\\
 		=& \mathbf{r}- \mathbf{b}_x \eta_n- \mathbf{b}_z \xi_n,
 	\end{aligned}
 \end{equation}
 where $\mathbf{r}\triangleq {\rm vec}(\mathbf{R})$, $\mathbf{b}_x\triangleq \mathbf{f}_n^T\otimes  \frac{\partial \mathbf{b}(x_n,z_n)}{\partial x_n}\in\mathbb{C}^{K^2\times 1}$, and $\mathbf{b}_z\triangleq \mathbf{f}_n^T\otimes  \frac{\partial \mathbf{b}(x_n,z_n)}{\partial z_n} \in\mathbb{C}^{K^2\times 1}$.
 
 The problem is converted to the projection problem. That is, $\eta_n$ and $\xi_n$ can be obtained by projecting $\mathbf{r}$ onto $\mathbf{b}_x$ and $\mathbf{b}_z$, respectively. However, the projection should be regularized for satisfying the constraint in problem (\ref{TR}). Hence,
 we obtain $ {\eta}_n=\frac{\mathbf{b}^H_x\mathbf{r}}{\mathbf{b}^H_x\mathbf{b}_x+\gamma}$ and ${\xi}_n=\frac{\mathbf{b}^H_z\mathbf{r}}{\mathbf{b}^H_z\mathbf{b}_z+\gamma}$, where $\gamma$ is the regularization factor which can be determined by the biselection method to satisfy the inter-element spacing constraint\textsuperscript{\ref {note1}}. Subsequently, 
 the updating rule for $x_n$ and $z_n$ is given by:
$
 	x_n\leftarrow x_n+\Re\{\eta_n\}
$,
$
 	z_n\leftarrow z_n+\Re\{\xi_n\},
$
 where $\Re\{\cdot\}$ is used for constraining the real coordinates.
  
 After updating the $n$-th antenna position, the channel and precoding can be updated, and then for its next iteration.
 As a result, the above deduction can provide us an iterative solution for ${x}_n$ and ${z}_n$ for	$\widehat{\bm{\mathcal{G}}}=\textbf{OffG\_Position\_Refinement} \left(
 \widehat{\bm{\mathcal{G}}},\bm{\theta},\bm{\phi}\right)$.
\end{appendices}

\bibliographystyle{IEEEtran}
\bibliography{reference.bib}

\vspace{12pt}

\end{document}